# Theory of Andreev reflection spectroscopy with anisotropic spin-dependent scattering*

LI Zhiyue[1, 2], ZHANG Guoping[1], CHEN Tingyong[1, 2, *]

1. Shenzhen Institute for Quantum Science and Engineering, Southern University of Science and Technology, Shenzhen 518055, China
2. International Quantum Academy, Shenzhen 518048, China

**Abstract**
Spintronic technologies require efficient generation and control of spin-polarized currents. Conventional ferromagnet-based methods suffer from sensitivity to external magnetic fields. Andreev reflection spectroscopy is vital for measuring spin polarization and superconducting gaps, yet prevailing theories assume isotropic interface scattering. This neglects ubiquitous anisotropy in heterostructures, causing misinterpretation of material properties. To resolve this, we develop a generalized model incorporating spin-dependent anisotropic scattering. Introducing distinct interface barriers for spin-up and spin-down electrons extends both the Blonder-Tinkham-Klapwijk formalism and the Chen-Tesanovic-Chien extension. This unified framework describes transport from normal metals to half-metals. Solving the Bogoliubov-de Gennes equations with modified boundary conditions yields current formulae with a transmission probability judgment function identifying dominant spin channels. In non-magnetic metals, interfacial anisotropy generates sizable spin-polarized currents via transmission spin filtering, suppressing Andreev reflection and reducing sub-gap conductance. For positively polarized ferromagnets, anisotropy nonlinearly modulates polarization, enhancing Andreev reflection to a threshold before suppression. Negatively polarized materials exhibit inverse spectra, enabling unambiguous polarization sign determination via conductance comparisons. Epitaxial Co film measurements validate the model, resolving subtle anisotropies. This refines Andreev spectrum interpretation and supports interference-resistant spin sources using non-magnetic platforms, benefiting magnetoresistive devices and superconducting

---

quantum technologies.

# 1 Introduction

The core challenge in spintronics lies in the efficient generation and manipulation of spin-polarized currents[1]. Traditionally, this objective has relied primarily on spin band splitting induced by exchange interactions in ferromagnetic materials[2-4]. However, spintronic devices based on magnetic materials exhibit significant limitations. In particular, their performance is susceptible to interference from external magnetic fields. This vulnerability has prompted researchers to explore new pathways for generating spin-polarized currents using non-magnetic materials[5-8]. In recent years, quantum materials such as topological insulators have been theoretically predicted to generate highly polarized currents due to their unique spin-momentum locking properties. Nevertheless, precise experimental measurement of the spin polarization in these materials remains challenging.

Among various characterization techniques, spectroscopic methods based on Andreev reflection at metal/superconductor interfaces demonstrate distinct advantages. These methods can simultaneously probe the spin polarization ($P$) and the superconducting energy gap ($\Delta$) of materials[9]. When electrons with energy below the superconducting gap incident from a metal onto a superconductor, they combine with electrons of opposite spin in the superconductor to form Cooper pairs. This process reflects holes back into the metal, leading to a significant enhancement in conductance. For non-magnetic metals, the normalized differential conductance at an ideal interface is approximately 2. In contrast, for half-metals, Andreev reflection is completely suppressed due to the absence of electrons with opposite spin, causing the conductance to approach zero.

The Blonder-Tinkham-Klapwijk (BTK) model[10] provides a theoretical foundation for the transport properties of normal metal/superconductor junctions. Strijkers et al.[11] extended the BTK model to describe transport in magnetic metal/superconductor junctions by introducing spin-dependent density of states parameters. However, their model did not explicitly account for the anisotropy of interface scattering. Subsequently, the Chen-Tesanovic-Chien (CTC) model[12] generalized this approach to magnetic metal systems by linearly combining fully polarized and unpolarized scenarios. Nevertheless, these traditional models assume that the interface exhibits identical scattering characteristics for electrons with different spin orientations, i.e., $Z_\uparrow = Z_\downarrow$. This assumption differs significantly from many actual interface systems. Experimental studies indicate that significant spin-dependent scattering often exists at interfaces in spintronic devices such as giant magnetoresistance (GMR) and tunnel magnetoresistance (TMR) devices[13-16]. For instance, early theoretical models by Mazin et al.[17] for half-metal/superconductor junctions demonstrated that Andreev reflection is completely suppressed when only one spin direction of electronic states exists at the Fermi level. Later, Wilken and Brouwer[18] observed in Fe/Cr interface experiments that anisotropic interface scattering can lead to pronounced spin filtering effects. More recently, Rigato et al.[3] discovered in ferromagnetic insulator/superconductor junctions that the spin dependence of the interface barrier can significantly modulate current polarization. These experimental results suggest that actual interfaces may present different scattering strengths for spin-up and spin-down electrons. Neglecting the anisotropy of interface scattering may lead to misinterpretation of experimental data, resulting in erroneous judgments of the intrinsic spin polarization of materials. Especially when investigating novel quantum materials, such deviations may obscure intrinsic material properties and even lead to incorrect physical conclusions. Therefore, developing theoretical models that fully account for anisotropic interface scattering effects is crucial not only for precisely characterizing material spin properties but also for guiding the design of novel spintronic devices.

This study aims to establish a theoretical framework for anisotropic spin scattering to systematically investigate the regulation of Andreev reflection spectra by interface scattering anisotropy. By introducing spin-dependent scattering parameters $Z_\uparrow$ and $Z_\downarrow$, this work reveals that anisotropic interfaces can induce highly polarized currents even in non-magnetic metals. This finding provides new insights for developing interference-resistant spintronic devices. Furthermore, this theory offers a foundational basis for explaining the modulation effects of interface engineering on the spin polarization of magnetic metals. These research outcomes will help promote the precise application of Andreev reflection spectroscopy in the characterization of novel quantum materials and the design of spintronic devices.

# 2 Physical Model and Theoretical Method

The theory of anisotropic spin scattering builds upon the classical BTK model by further considering the physical scenario where the metal/superconductor interface exhibits different scattering strengths for spin-up and spin-down electrons. This model can uniformly describe the charge transport properties at interfaces between various magnetic metals, ranging from normal metals ($P$ = 0) to half-metals ($P$ = ±1), and s-wave superconductors.

## 2.1 BTK Model

Based on the BCS theory framework for s-wave superconductors, the Hamiltonian of the system can be expressed as

$$H_{\mathrm{BCS}} = \sum_{v>0} \varepsilon_v (a_v^+ a_v + a_{-v}^+ a_{-v}) - V \sum_{\mu>0, v>0} a_\mu^+ a_{-\mu}^+ a_{-v} a_v, \tag{1}$$

Here, $\varepsilon_v$ denotes the single-electron kinetic energy, $a_v^+$ and $a_v$ represent the electron creation and annihilation operators, respectively, and $V$ signifies the effective attractive potential. By introducing the order parameter $\Delta = V \sum_v \langle a_{-v} a_v \rangle$ through the mean-field approximation, we obtain the simplified form:

$$H_{\mathrm{BCS}} = \sum_v \varepsilon_v (a_v^+ a_v + a_{-v}^+ a_{-v}) + \frac{1}{2} \sum_v (\Delta_v a_v^+ a_{-v}^+ + \Delta_v^* a_{-v} a_v). \tag{2}$$

This Hamiltonian describes the formation of Cooper pairs in superconductors and serves as the theoretical foundation for analyzing interfacial Andreev reflection.

The BTK model is the standard theory for treating electron transmission across normal metal/superconductor (N/S) interfaces. Based on the Bogoliubov equations[19,20], this model assumes a scattering barrier at the interface in the form of a delta function, $V(x) = H\delta(x)$, where $H$ denotes the interfacial scattering strength. The continuity and jump conditions for the wave function at the interface can be expressed as

$$\psi_{\mathrm{S}}(0) = \psi_{\mathrm{N}}(0), \ \frac{\hbar^2}{2m}(\psi_{\mathrm{S}}'(0) - \psi_{\mathrm{N}}'(0)) = H\psi(0). \tag{3}$$

The incident, reflected, and transmitted wave components are expressed as follows:

$$\begin{aligned} \psi_{\mathrm{inc}} &= \binom{1}{0} \mathrm{e}^{\mathrm{i}q^+ x}, \\ \psi_{\mathrm{refl}} &= a\binom{0}{1} \mathrm{e}^{\mathrm{i}q^- x} + b\binom{1}{0} \mathrm{e}^{-\mathrm{i}q^+ x}, \\ \psi_{\mathrm{trans}} &= c\binom{u_0}{v_0} \mathrm{e}^{\mathrm{i}k^+ x} + d\binom{v_0}{u_0} \mathrm{e}^{-\mathrm{i}k^- x}, \end{aligned} \tag{4}$$

Here, $\hbar q^{\pm} = \sqrt{2m}(\mu \pm E)^{1/2}$ represents the momentum of incident normal electrons, while $\hbar k^{\pm} = \sqrt{2m}[\mu \pm (E^2 - \Delta^2)^{1/2}]^{1/2}$ denotes the quasiparticle momentum in the superconductor. The coefficients $u_0$ and $v_0$ satisfy $u_0^2 = 1 - v_0^2 = \frac{1}{2} \times [1 + \frac{(E^2 - \Delta^2)^{1/2}}{E}]$. Under the weak energy approximation ($E \ll \mu$), these simplify to $q^+ = q^- = k^+ = k^- = k_\mathrm{F}$. Here, $u_0$ and $v_0$ are the Bogoliubov coefficients, and $k_\mathrm{F}$ is the Fermi wave vector.

On the metallic side ($x < 0$), the wave function comprises a superposition of the incident electron wave, the normally reflected electron wave, and the Andreev reflected hole wave:

$$\psi_\mathrm{N}(x) = \binom{1}{0}\mathrm{e}^{\mathrm{i}q_\mathrm{e}^+ x} + b(E)\binom{1}{0}\mathrm{e}^{-\mathrm{i}q_\mathrm{e}^+ x} + a(E)\binom{0}{1}\mathrm{e}^{\mathrm{i}q_\mathrm{h}^- x}, x < 0. \tag{5}$$

On the superconducting side ($x > 0$), the wave function is a superposition of transmitted electron-like and hole-like quasiparticle waves:

$$\psi_\mathrm{S}(x) = c(E)\binom{u_0}{v_0}\mathrm{e}^{\mathrm{i}k_\mathrm{e}^+ x} + d(E)\binom{v_0}{u_0}\mathrm{e}^{-\mathrm{i}k_\mathrm{h}^- x}, \quad x > 0. \tag{6}$$

Substituting the wave function into the boundary conditions yields the four probability amplitude coefficients.

When $E > \Delta$, the reflection and transmission coefficients are

$$a = \frac{u_0 v_0}{\gamma}, b = -\frac{(u_0^2 - v_0^2)(Z^2 + \mathrm{i}Z)}{\gamma}, \quad c = \frac{u_0(1 - \mathrm{i}Z)}{\gamma}, d = \frac{\mathrm{i}v_0 Z}{\gamma}, \tag{7}$$

Here, $Z = mH/\hbar^2 k_\mathrm{F} = H/\hbar v_\mathrm{F}$ denotes the dimensionless interface scattering parameter, and $\gamma = u_0^2 + (u_0 - v_0^2)Z^2$. The corresponding probability current is

$$A = \frac{J_\mathrm{P}^A}{v_\mathrm{F}} = a^* a, B = \frac{J_\mathrm{P}^B}{v_\mathrm{F}} = b^* b, \quad C = \frac{J_\mathrm{P}^C}{v_\mathrm{F}} = c^* c(u_0^2 - v_0^2), D = \frac{J_\mathrm{P}^D}{v_\mathrm{F}} = d^* d(u_0^2 - v_0^2). \tag{8}$$

When $E < \Delta$[17], $k^\pm$, $u_0$, and $v_0$ are all complex numbers. The Andreev reflection probability is given by

$$A = \frac{\Delta^2}{E^2 + (\Delta^2 - E^2)(1 + 2Z^2)^2}, B = 1 - A. \tag{9}$$

At this point, the normal transmission probabilities $C = D = 0$.

The BTK model reveals enhanced conductance within the energy gap due to Andreev reflection and conductance approaching normal values outside the gap, laying the foundation for subsequent models[21,22].

## 2.2 Half-metal Model

The half-metal model (Mazin model) addresses metal/superconductor interfaces with complete spin polarization ($P = \pm 1$)[12,17]. Since only electrons of a single spin direction exist at the Fermi level, Andreev reflection is completely suppressed because the formation of Cooper pairs in the superconductor requires electrons with opposite spins. When $E > \Delta$, the reflection and transmission coefficients are

$$\begin{gathered} a = 0, b = -\frac{-u^2 Z + v^2(\mathrm{i}+z)}{\gamma_2}, \\ c = -\frac{\mathrm{i}uZ}{\gamma_2}, d = \frac{\mathrm{i}uZ}{\gamma_2}, \end{gathered} \tag{10}$$

Here, $\gamma_2 = u^2(\mathrm{i} + z) - v^2 Z^2$. The corresponding probability current is

$$\begin{gathered} A = 0, B = \frac{J_{\mathrm{P}}^{B}}{v_{\mathrm{F}}} = b^* b, \\ C = \frac{J_{\mathrm{P}}^{C}}{v_{\mathrm{F}}} = c^* c(u_0^2 - v_0^2), \\ D = \frac{J_{\mathrm{P}}^{D}}{v_{\mathrm{F}}} = d^* d(u_0^2 - v_0^2). \end{gathered} \tag{11}$$

When $E$ $\Delta$, the absence of Andreev reflection and electron transmission results in total electron reflection:

$$a = 0, b = 1, c = 0, d = 0. \tag{12}$$

The probability current is

$$A = 0, B = 1, C = 0, D = 0. \tag{13}$$

The half-metallic model indicates that electrical conductance within the energy gap is zero under conditions of complete spin polarization. This stands in sharp contrast to the BTK model. The modified BTK model (CTC model) linearly combines the BTK model with the half-metallic model to address interfaces involving general ferromagnetic metals ($0 < P < 1$).

## 2.3 Derivation of Current Formula under Anisotropic Interfaces

In ferromagnetic metal/superconductor interfaces, the transport of spin-polarized current depends not only on the intrinsic spin polarization ($P$) of the material but also significantly on the anisotropy of interface scattering. Traditional models, such as the CTC model, assume isotropic interface scattering (i.e., $Z_\uparrow = Z_\downarrow$). They describe general ferromagnetic metals by linearly combining the limits of non-polarized ($P = 0$) and fully polarized ($P = \pm 1$) cases. However, realistic interfaces, such as oxide layers or magnetic impurities, often exhibit different scattering strengths for spin-up and spin-down electrons (i.e., $Z_\uparrow \neq Z_\downarrow$). This anisotropy causes an imbalance in the number of electrons transmitted into the superconductor between spin channels, thereby affecting the efficiency of Andreev reflection. Andreev reflection requires the pairing of electrons with opposite spins to form Cooper pairs. Consequently, its probability depends on the comparison of transmission rates between the two spin channels, rather than merely the weighted average of incident electron numbers. If the transmission rate of one spin channel is significantly higher than that of the other, Andreev reflection becomes limited by the channel with the lower transmission rate, known as the "bottleneck effect." This aligns with the physics of spin-dependent scattering in GMR/TMR devices[1]. This section extends the modified BTK model by introducing spin-dependent scattering parameters $Z_\uparrow$ and $Z_\downarrow$, which represent the interface scattering strengths for spin-up and spin-down electrons, respectively.

The spin polarization of metallic materials is defined as

$$P \equiv \frac{N_\uparrow(0)v_{\mathrm{F}\uparrow} - N_\downarrow(0)v_{\mathrm{F}\downarrow}}{N_\uparrow(0)v_{\mathrm{F}\uparrow} + N_\downarrow(0)v_{\mathrm{F}\downarrow}}, \tag{14}$$

Here, $N_\uparrow(0)$ and $N_\downarrow(0)$ denote the spin-up and spin-down densities of states at the Fermi level, respectively, while $v_{\mathrm{F}\uparrow}$ and $v_{\mathrm{F}\downarrow}$ represent the corresponding Fermi velocities. The average density of states and Fermi velocity are defined as

$$\begin{aligned} 2N(0) &\equiv N_\uparrow(0) + N_\downarrow(0), \\ v_\mathrm{F} &\equiv \frac{N_\uparrow(0)v_{\mathrm{F}\uparrow} + N_\downarrow(0)v_{\mathrm{F}\downarrow}}{N_\uparrow(0)+N_\downarrow(0)}. \end{aligned} \tag{15}$$

Thus, we obtain

$$N_\uparrow(0)v_{\mathrm{F}\uparrow} = (1+P)N(0)v_\mathrm{F}, \tag{16}$$

$$N_\downarrow(0)v_{\mathrm{F}\downarrow} = (1-P)N(0)v_\mathrm{F}. \tag{17}$$

The current formula for the magnetic metal/superconductor interface must treat spin-up and spin-down electrons separately. The total current is

$$\begin{aligned} I_\mathrm{MS} = {} & N_\uparrow(0)ev_{\mathrm{F}\uparrow}S\int_{-\infty}^{\infty} [f_0(E-eV) - f_0(E)] \\ & \times [1 + A_\uparrow(E) - B_\uparrow(E)]\mathrm{d}E \\ & + N_\downarrow(0)ev_{\mathrm{F}\downarrow}S\int_{-\infty}^{\infty} [f_0(E-eV) - f_0(E)] \\ & \times [1 + A_\downarrow(E) - B_\downarrow(E)]\mathrm{d}E, \end{aligned} \tag{18}$$

Here, $e$ denotes the elementary charge, and $S$ represents the effective interfacial cross-sectional area. $f_0(E) = (\mathrm{e}^{E/k_\mathrm{B}T} + 1)^{-1}$ is the Fermi-Dirac distribution function, where $k_B$ is the Boltzmann constant and $T$ is the temperature. $A_\sigma(E)$ and $B_\sigma(E)$ denote the probabilities of Andreev reflection and normal reflection, respectively, for spin $\sigma$ ($\uparrow$ or $\downarrow$). These probabilities are calculated using the BTK model (with $Z_\sigma$ replacing $Z$). Substituting definitions (16) and (17) into equation (8) yields

$$I_{MS} = N(0)\mathrm{e}v_F S\int_{-\infty}^{\infty} [f_0(E-\mathrm{e}V)-f_0(E)] \times \{(1+P)[1+A_\uparrow(E)-B_\uparrow(E)] + (1-P)[1+A_\downarrow(E)-B_\downarrow(E)]\}\mathrm{d}E. \quad (19)$$

The key innovation lies in introducing a judgment condition to compare the transmission probabilities of spin-up and spin-down electrons, expressed as $1 + A_\sigma(E) - B_\sigma(E)$. We define the judgment function as follows:

Spin-up electrons dominate when $(1+P)[1+A_\uparrow(E)-B_\uparrow(E)] \geqslant (1-P)[1+A_\downarrow(E)-B_\downarrow(E)]$.

$$(1+P)[1+A_\uparrow(E)-B_\uparrow(E)] < (1-P)[1+A_\downarrow(E)-B_\downarrow(E)]$$

At this time, spin-down electrons dominate.

The current formula is expressed piecewise based on the judgment criteria as follows.

When spin-up electrons dominate:

$$I_{MS} = N(0)\mathrm{e}v_F S\int_{-\infty}^{\infty} [f_0(E-\mathrm{e}V)-f_0(E)]\{2(1-P)[1+A_\downarrow(E)-B_\downarrow(E)] + \frac{(1+P)[1+A_\uparrow(E)-B_\uparrow(E)]-(1-P)[1+A_\downarrow(E)-B_\downarrow(E)]}{(1+P)[1+A_\uparrow(E)-B_\uparrow(E)]}[1-B_p(E)]\}\mathrm{d}E, \quad (20)$$

Here, $B_p(E)$ denotes the normal reflection probability in the half-metal model, adopted from the Mazin model to characterize the reflection behavior of fully polarized current. When interfacial anisotropy is strong, transmission through minor channels becomes negligible, and the system reduces to a half-metal-superconductor junction. This treatment aligns with experimental observations, such as the spin-filtering effect in ferromagnet/superconductor junctions[4]. We calculate using $Z = Z_\uparrow$ (i.e., the scattering parameter for spin-up electrons).

When spin-down electrons dominate,

$$I_{MS} = N(0)\mathrm{e}v_F S\int_{-\infty}^{\infty} [f_0(E-\mathrm{e}V)-f_0(E)]\{2(1-P)[1+A_\uparrow(E)-B_\uparrow(E)] + \frac{(1+P)[1+A_\downarrow(E)-B_\downarrow(E)]-(1-P)[1+A_\uparrow(E)-B_\uparrow(E)]}{(1+P)[1+A_\downarrow(E)-B_\downarrow(E)]}[1-B_p(E)]\}\mathrm{d}E, \quad (21)$$

Here, $B_{\mathrm{p}}(E)$ is calculated using $Z = Z_{\downarrow}$.

This piecewise approach ensures that current calculations align more closely with physical reality. When one channel dominates, system behavior approaches the half-metallic model, where Andreev reflection is suppressed. When transmission is balanced, the system reverts to the BTK model. This strategy avoids deviations that traditional linear combinations may produce at highly anisotropic interfaces.

To facilitate experimental comparison, we introduce the normalized resistance $R_{\mathrm{N}}$. When the bias voltage $V$ satisfies $\mathrm{e}V \gg \Delta$, the current simplifies to

$$I_{\mathrm{MS}} = [\frac{1-P}{1+Z_{\downarrow}^{2}} + \frac{1+P}{1+Z_{\uparrow}^{2}}]N(0)\mathrm{e}^{2}v_{\mathrm{F}}SV \equiv \frac{V}{R_{\mathrm{N}}}, \quad (22)$$

Solving yields

$$N(0)\mathrm{e}v_{\mathrm{F}}S = \frac{1}{\mathrm{e}R_{\mathrm{N}}}[\frac{1-P}{1+Z_{\downarrow}^{2}} + \frac{1+P}{1+Z_{\uparrow}^{2}}]^{-1}. \quad (23)$$

Substituting into Equation (20) or (21) yields the final $I$-$V$ relationship, which can be used to fit experimental data. This model unifies the limiting cases of the BTK model ($P = 0$, $Z_{\uparrow} = Z_{\downarrow}$) and the half-metal model ($P = \pm1$, $Z_{\uparrow} = Z_{\downarrow}$). Systematic comparison clarifies its novelty. The BTK model[5] assumes isotropic interface scattering (i.e., $Z_{\uparrow} = Z_{\downarrow}$). It applies to non-magnetic metals ($P = 0$) but fails to describe spin-dependent scattering effects. The Mazin model[17] further addresses the half-metal case ($P = \pm1$). However, it is limited to fully polarized interfaces and ignores the transitional behavior of partially polarized materials. The CTC model[12], as a linear combination of the BTK and Mazin models, retains the isotropy assumption. Although it can describe general magnetic metals ($0<P<1$), it does not take into account the common scattering anisotropy in actual interfaces The Strijkers model focuses on weighting the density of states of bulk materials, emphasizing intrinsic polarizability, but does not explicitly address the asymmetry of interface scattering This article introduces spin dependent parameters and a judgment function to unify the above limit cases: when $Z_{\uparrow} = Z_{\downarrow}$, the model degenerates into CTC form; Under strong anisotropy conditions (e.g., $Z_{\uparrow} \gg Z_{\downarrow}$), the system approaches Mazin behavior. Furthermore, it captures the modulation of polarization by interface engineering, complementing the Strijkers model. This generalization not only refines the theoretical description but also provides new insights for explaining interface enhancement effects, such as TMR/GMR optimization.

# 3 Anisotropic Andreev Reflection Spectra

Based on the anisotropic spin scattering theory model, this section conducts in-depth research on the regulation of interface scattering anisotropy on Andreev reflection spectra under different spin polarization conditions through systematic numerical calculations Focus on analyzing the evolution characteristics of normalized differential conductance under three scenarios of P=0, P>0, and P<0, and explore their physical mechanisms For the convenience of image comparison, the usually drawn image is a normalized dI/dV image. Therefore, we take the derivative of V on both sides of the final I-V relationship, and then divide it by the normalized resistance RN on both sides of the formula to obtain the following two formulas:

$$\frac{1}{R_N}\frac{dI_{MS}}{dV} = \left[\frac{1}{\frac{1-P}{1+Z_\downarrow^2}+\frac{1+P}{1+Z_\uparrow^2}}\right]\int_{-\infty}^{\infty}\frac{\exp(\frac{E-eV}{k_BT})}{[\exp(\frac{E-eV}{k_BT})+1]^2\cdot k_BT}\{2(1-P)[1+A_\downarrow(E)-B_\downarrow(E)] + \frac{(1+P)[1+A_\uparrow(E)-B_\uparrow(E)]-(1-P)[1+A_\downarrow(E)-B_\downarrow(E)]}{(1+P)[1+A_\uparrow(E)-B_\uparrow(E)]}[1-B_p(E)]\}dE, \tag{24}$$

$$\frac{1}{R_N}\frac{dI_{MS}}{dV} = \left[\frac{1}{\frac{1-P}{1+Z_\downarrow^2}+\frac{1+P}{1+Z_\uparrow^2}}\right]\int_{-\infty}^{\infty}\frac{\exp(\frac{E-eV}{k_BT})}{[\exp(\frac{E-eV}{k_BT})+1]^2\cdot k_BT}\{2(1+P)[1+A_\uparrow(E)-B_\uparrow(E)] + \frac{(1-P)[1+A_\downarrow(E)-B_\downarrow(E)]-(1+P)[1+A_\uparrow(E)-B_\uparrow(E)]}{(1-P)[1+A_\downarrow(E)-B_\downarrow(E)]}[1-B_p(E)]\}dE. \tag{25}$$

## 3.1 Andreev Reflection Spectra at $P = 0$

In the case of a normal metal ($P = 0$), the Andreev reflection spectra at an anisotropic spin-scattering interface exhibit distinct characteristics. Figure 1 presents the normalized differential conductance when the interface scattering coefficients for spin-up and spin-down electrons ($Z_\uparrow$ and $Z_\downarrow$) vary symmetrically. When $Z_\uparrow = Z_\downarrow = 0$, the normalized differential conductance equals 2 within the energy gap ($|E| \leq \Delta$). This aligns with the classical BTK model, indicating that Andreev reflection dominates. However, as $Z_\uparrow$ increases while $Z_\downarrow$ remains fixed at 0, the differential conductance within the gap gradually decreases to zero. This occurs because the reduced transmission of spin-up electrons hinders Cooper pair formation, thereby suppressing Andreev reflection[23]. Outside the energy gap ($|E| > \Delta$), the differential conductance approaches 1, suggesting that normal reflection becomes dominant.

To further investigate the behavior at small $Z$ values, Figure 2 presents the magnified spectra of $Z_\uparrow$ and $Z_\downarrow$ within the range of 0 to 1. The differential conductance exhibits sensitivity to variations in $Z$. Even small $Z$ values, such as 0.25 or 0.5, significantly affect the Andreev reflection efficiency. This finding underscores the importance of precise control over interface scattering.

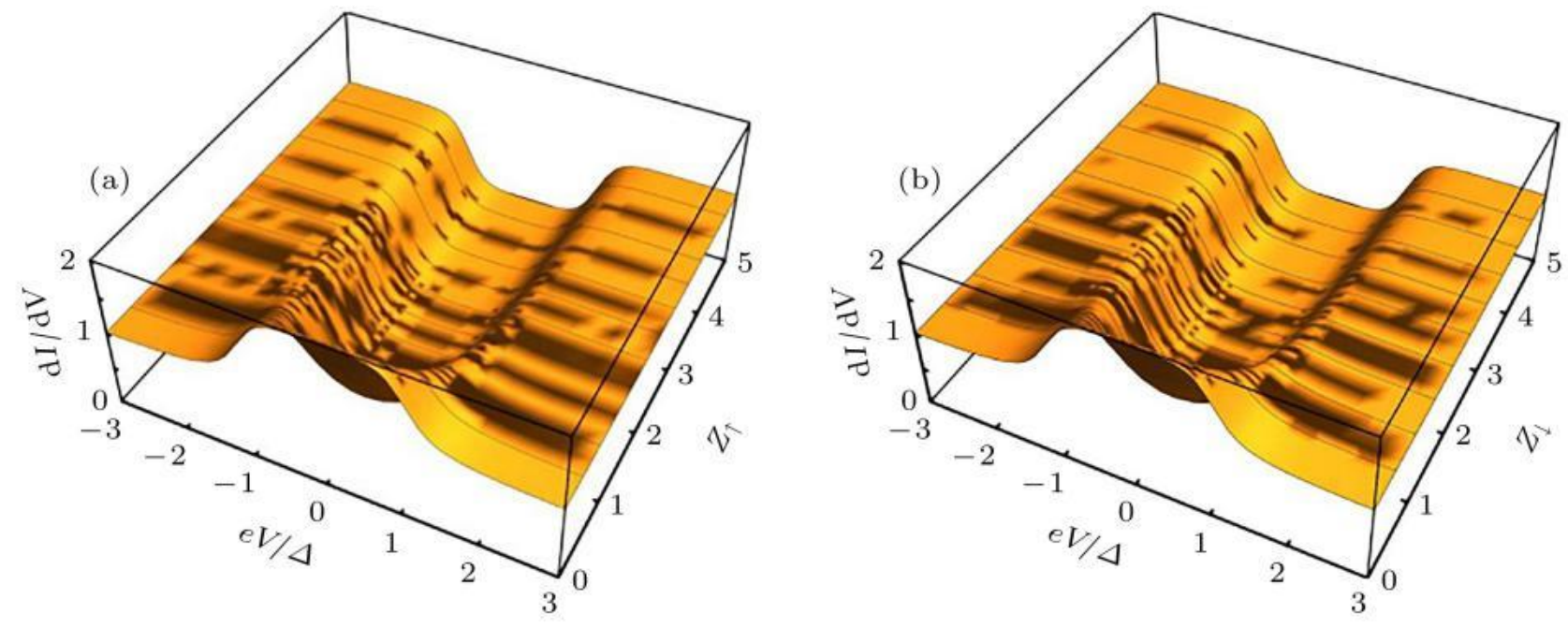

**Fig. 1 Normalized differential conductance at symmetric anisotropic interfaces for $P$ = 0: (a) $Z_{\downarrow}$= 0, $Z_{\uparrow}$= 0-5; (b) $Z_{\uparrow}$= 0, $Z_{\downarrow}$= 0-5.**

The results above indicate that even for nonmagnetic metals ($P$ = 0), anisotropic scattering at the interface can induce highly spin-polarized currents outside the energy gap. This finding offers potential for spintronic devices based on nonmagnetic materials.

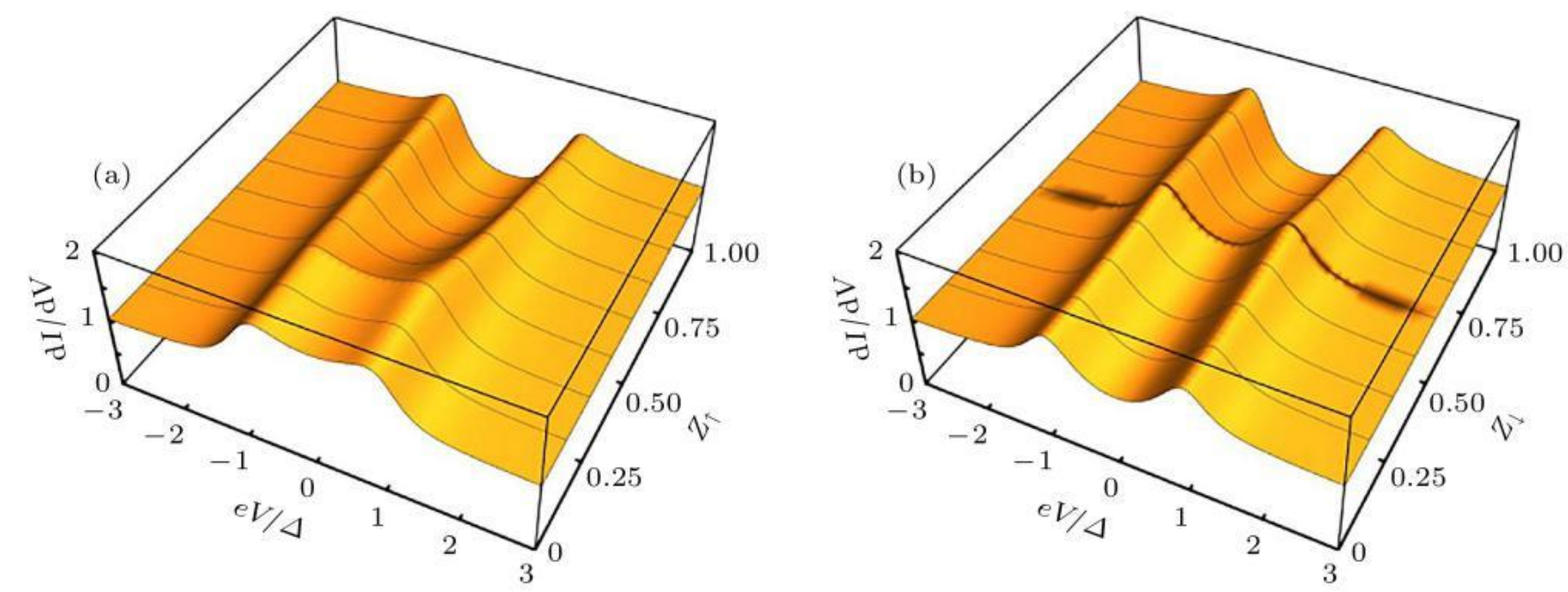

**Fig. 2 Normalized differential conductance at symmetric anisotropic interfaces for a nonmagnetic metal ($P$ = 0): (a) $Z_{\uparrow}$= 0.25, $Z_{\downarrow}$= 0-1; (b) $Z_{\uparrow}$= 0.5, $Z_{\downarrow}$= 0-1.**

## 3.2 Andreev Reflection Spectra for $P$ > 0

When the metal exhibits positive spin polarization ($P$ > 0), the Andreev reflection spectra at anisotropic interfaces display nonlinear characteristics (Figure 3). Figure 3(a) illustrates the case where $P$ = 0.25, with $Z_{\downarrow}$ fixed at 0.5 and $Z_{\uparrow}$ increasing from 0 to 5. When $Z_{\uparrow}$< $Z_{\downarrow}$, the differential conductance rises as $Z_{\uparrow}$ increases. This occurs because spin-up electrons are more abundant, and increasing $Z_{\uparrow}$ reduces normal reflection of these excess electrons while enhancing Andreev reflection. However, once $Z_{\uparrow}$ exceeds a critical value (approximately 0.68), the differential conductance decreases. This decline indicates that spin-down electrons become dominant, making Cooper pair formation more difficult. In contrast, Figure 3(e) shows that when $Z_{\uparrow}$ is fixed at 0.5 and $Z_{\downarrow}$ increases, the differential conductance decreases monotonically without any peak. This behavior arises because spin-up electrons dominate when $P$ > 0, so increasing $Z_{\downarrow}$ directly suppresses the contribution from the minority spin channel.

As $P$ increases, the effect becomes more pronounced. When $P$ gradually rises to 0.5 and 0.75, the peak of the differential conductance shifts toward larger $Z_\uparrow$ values, as stronger scattering is required to balance the spin distribution. At $P$ = 0.95, the spectrum approaches the half-metallic model, with Andreev reflection almost completely suppressed. These results indicate that interfacial anisotropy can modulate current polarization, causing it to deviate from the intrinsic $P$ value of the material. This provides a theoretical basis for explaining interface enhancement effects in devices such as TMR or GMR.

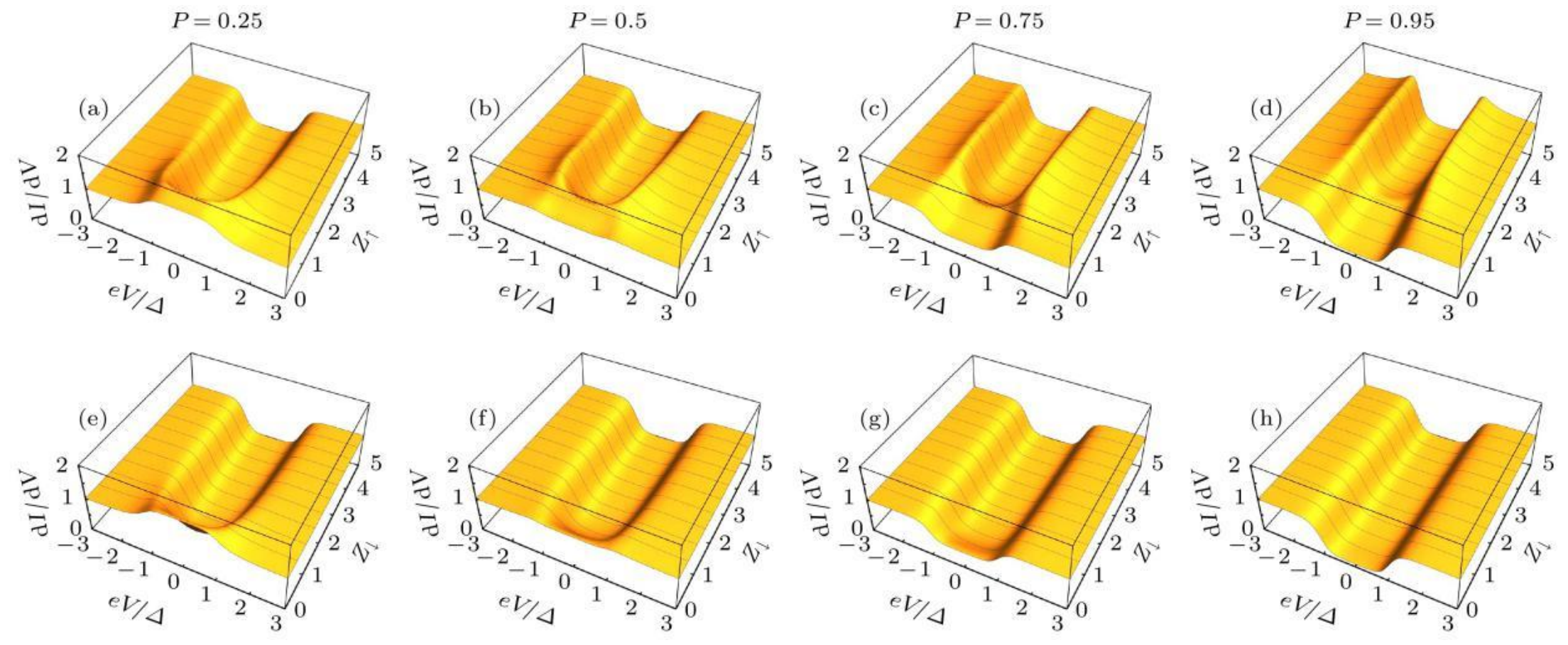

**Fig. 3 Normalized differential conductance at anisotropic interfaces for a magnetic metal (*P* > 0): (a)-(d) $Z_\downarrow$= 0.5, $Z_\uparrow$= 0-5; (e)-(h) $Z_\uparrow$= 0.5, $Z_\downarrow$= 0-5.**

### 3.3 Andreev Reflection Spectra for $P < 0$

When $P < 0$, the anisotropic Andreev reflectance spectrum exhibits significantly different characteristics from $P > 0$, providing a new method for determining the sign of spin polarization Figure 4 (a) shows the case of $P$ = -0.25, where $Z_\downarrow$ is fixed at 0.5 and $Z_\uparrow$ increases from 0 to 5. Compared with $P$ = 0.25, Figure 4 (a) shows no differential conductivity peak because spin down electrons dominate when $P < 0$, and an increase in $Z_\uparrow$ directly suppresses Andreev reflection; Figure 4 (e) shows a peak, verifying the theoretical consistency.

As the $P$ value decreases (e.g., $P$ = -0.5, -0.75, and -0.95), the differential conductance spectra gradually flatten, approaching the half-metallic model. Specifically, in Fig. 4(f) and (h), when $Z_\uparrow$ is fixed, an increase in $Z_\downarrow$ leads to a peak in the differential conductance. The position of this peak shifts toward larger $Z_\downarrow$ values as $P$ decreases. This indicates that the sign of $P$ can be uniquely determined by measuring the Andreev reflection spectra at anisotropic interfaces. This approach provides a new pathway for resolving signal reversal issues in spintronic devices.

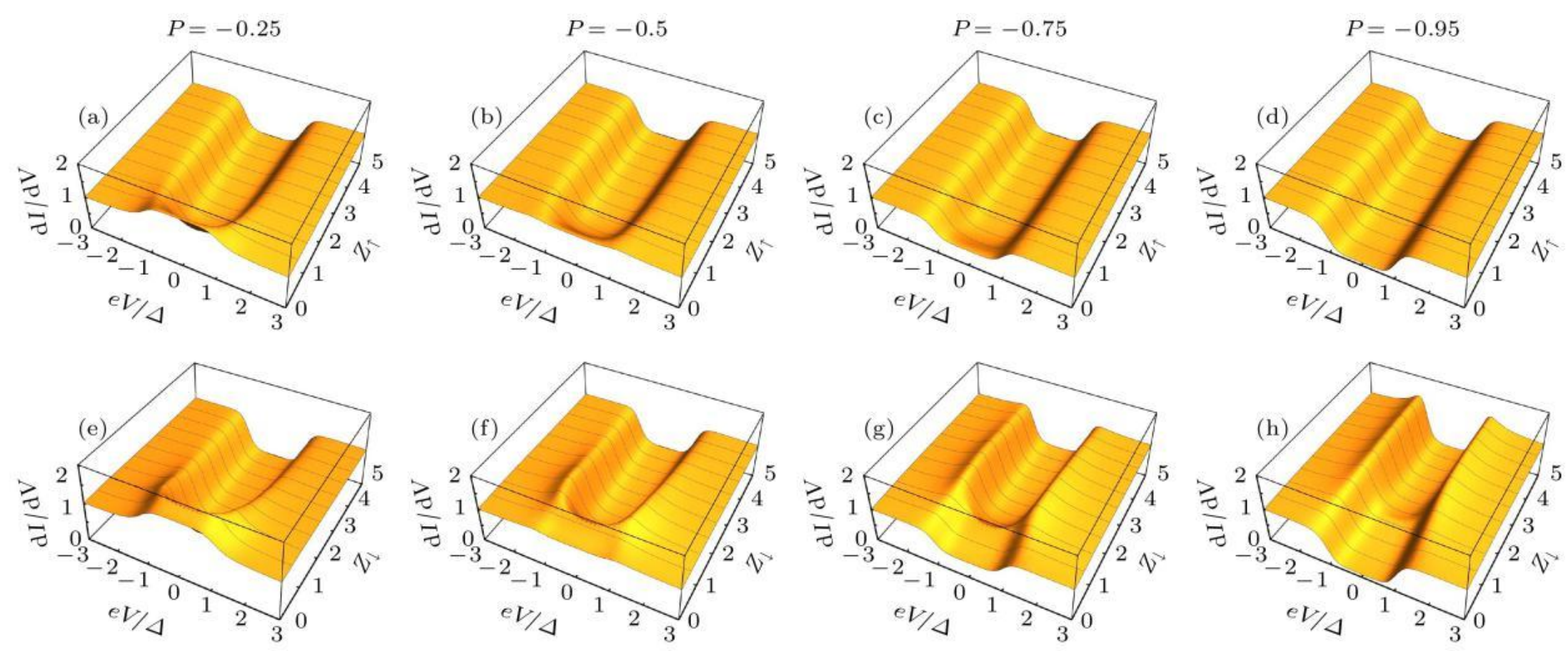

**Fig. 4 Normalized differential conductance at anisotropic interfaces for a negatively polarized material (*P* $Z_{\downarrow}$= 0.5, $Z_{\uparrow}$= 0-5; (e)-(h) $Z_{\uparrow}$= 0.5, $Z_{\downarrow}$= 0-5.**

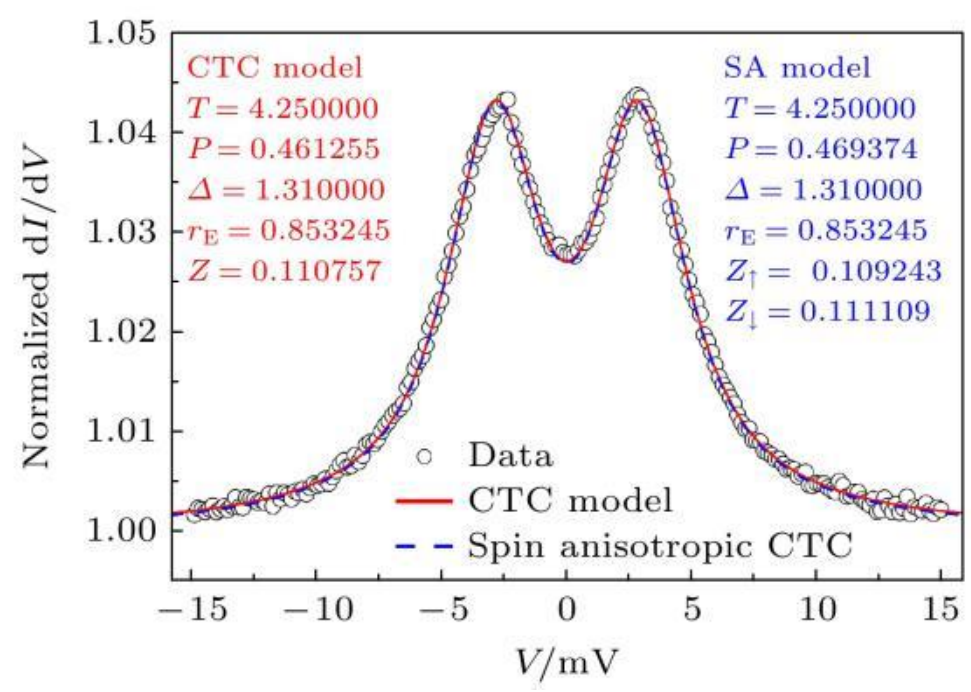

**Fig. 5 Andreev reflection spectroscopy measurements and model comparison for a pure Co film.**

To test the model, we prepared 50 nm Co films using magnetron sputtering and performed Andreev reflection spectroscopy using a superconducting Pb tip. We then fitted the data using both the spin-anisotropic scattering model proposed in this study and the CTC model, as shown in Fig. 5. The results indicate that spin-anisotropic scattering is minimal in pure Co films. The fitting results closely align with the CTC model, yielding a spin polarization consistent with typical values for Co[24]. Given the sample size of approximately 20 mm, the film resistance is relatively high, necessitating the introduction of an additional resistance $r_E$. In the spin-anisotropic scattering model, an increase in $Z_{\downarrow}$ leads to higher spin polarization. This implies that the spin polarization of Co is positive, which agrees with experimental observations.

# 4 Conclusion

Building upon the traditional BTK and CTC models, this study incorporates the anisotropic scattering effects prevalent at realistic interfaces. By introducing spin-dependent scattering parameters $Z_{\uparrow}$ and $Z_{\downarrow}$, we refine the theoretical description of Andreev reflection at metal/superconductor interfaces. For normal metals with a spin polarization rate of $P = 0$, interfacial anisotropic scattering ($Z_{\uparrow} \neq Z_{\downarrow}$) generates significant spin-polarized currents through a transport spin polarization mechanism. This finding offers new perspectives for developing spintronic devices based on non-magnetic materials. In magnetic metal systems, interfacial anisotropy effectively modulates current polarization. Furthermore, by systematically analyzing the distinct features of reflection spectra under conditions where $P > 0$ and $P < 0$, we propose a novel method to determine the sign of the material spin polarization rate. This approach provides new theoretical support for quantum material characterization and spintronic device design. Future research will focus on experimental validation and expanding the theoretical framework for low-dimensional material interfaces. Additionally, we will combine advanced algorithms to optimize parameter analysis, thereby promoting innovative applications of this theory in quantum information technology.